\documentclass[fleqn,usenatbib]{mnras}

\usepackage[T1]{fontenc}
\usepackage{ae,aecompl}
\usepackage{times}
\usepackage{color}
\usepackage{soul}

\usepackage{graphicx}	
\usepackage{amsmath}	
\usepackage{amssymb}	

\title[GW from long GRBs]{Gravitational wave bursts from long gamma-ray bursts}

\author[Nathanail \& De Laurentis]
{Antonios Nathanail$^{1}$\thanks{E-mail: nathanail@th.physik.uni-frankfurt.de}, Mariafelicia De Laurentis$^{1}$
\\
$^{1}$Institut f\"ur Theoretische Physik, Goethe Universit\"at Frankfurt,
Max-von-Laue-Str.1, 60438 Frankfurt am Main, Germany\\
}

\pubyear{2017}

\begin{document}
\label{firstpage}
\pagerange{\pageref{firstpage}--\pageref{lastpage}}
\maketitle

\begin{abstract}

One of the most luminous explosions detected, gamma-ray bursts, especially 
the so-called long-duration bursts, most probably consist
of an intrinsic core-collapse to a black hole 
inside a super massive star. We point out that this collapse 
alone will give a generic gravitational wave burst. It has been 
shown that the strength of this  burst depends on the dimensionless 
spin parameter of the collapsing object.  Under 
descent assumptions  the gamma-ray burst's central engine  powers the explosion 
electromagnetically 
due to the rotation of the newly formed black hole. We argue that 
the peak luminosity and the isotropic energy of the gamma-ray burst
can be associated  with the spin of the black hole, due 
to this mechanism. Since, both gravitational and electromagnetic emission 
depend on the spin, they can be correlated   and thus 
give a straight estimate for the gravitational wave burst, when we have in hand 
a gamma-ray burst with known distance. We discuss detectability 
limits for present and future detectors.

\end{abstract}




\section{Introduction}
\label{sec:intro}

%
The recent detection of gravitational waves from binary
black holes and binary neutron stars has signaled the beginning of the
gravitational wave astronomy and the multi-messenger era. 
The capabilities of present and future 
interferometric detectors have to be compared and 
correlated with all possible sources of gravitational waves (GW).
In this work we discuss a generic feature of the gravitational 
collapse to a black hole which is believed to play a fundamental 
role in the physical production of a 
long-duration gamma-ray bursts (hereafter GRBs).

Since their first discovery back in the $60$s \citep{Klebesadel1973}
GRBs have been a 
great scientific challenge both observationally
and, mostly, theoretically. These cosmological 
hyper-energetic explosions are likely also to 
release energy in the form of gravitational waves. 
The main mechanisms for producing them are compact stellar 
mergers, for short-duration GRBs and massive stellar 
collapses for long GRBs. The long and short characteristic 
was set up depending on the difference in their duration 
and through the years has led to these different  
mechanism characterization \citep{Kouveliotou1993}.

The connection between short GRBs and the 
coalescence of binary neutron stars was recently proven \citep{Abbott2017d}. 
The GW 
signal that will accompany such mergers, was the subject of 
extensive research both analytically and numerically
 \citep{Baiotti:2016rmd}.
On the other hand, for long GRBs there are studies 
discussing the GW production 
through different possible mechanisms, but the overall 
picture is subtle. 

The death of massive stars of mass  above 
$10 M_{\odot}$ is widely accepted  to originate a GRB. 
The main mechanism  supposes that the inner core of the massive star
directly collapses to a black hole. 
Other possibilities have been discussed such as the birth of a magnetar \citep{Usov:1992zd}. 
In this work we will stick with the previous 
assumption of a direct collapse to a black hole. 

During the death of the massive star, 
its core will pass from a neutron star phase and then 
collapse to a black.  The "collapsar" engine argues 
that hyper accretion from the stellar envelopes that have 
formed an accretion disk and/or energy extraction from 
the newly formed black hole will drive a 
relativistic jet, which will power the GRB \citep{Paczynski:1998qr,MacFadyen1999,Lee2000}.

In this context, there have been discussed some possible 
ways of GW production.
When the black hole is formed, after the collapse of the inner core 
of a massive star, accretion of the remaining stellar shells will distort 
the black hole resulting in a well-studied "ring-down" \citep{Berti:2014bla} dissipating all this accretion-induced excitation till the black hole settles down.

Another  possibility discussed in the literature is 
that during the collapse of the inner core, 
instabilities will result to the development of fragmentation and the production 
of multiple compact components whose coalescence will be a powerful 
emitter of GW \citep{Nakamura1989,Bonnell1995,vanPutten:2001gi}. 

The production of an accretion disk is also expected in the 
context of the core collapse of a massive star. 
If the mass of this disk is high enough for its self-gravity to be 
important, then gravitational instabilities could result 
in bar formation which would radiate GW \citep{Fryer:2002ji,vanPutten2002ApJ,Davies:2002fb}.
Following the formation of a puffed up accretion disk,  the
 possibility of electromagnetic interaction between 
 the black hole and  the torus could give rise to gravitational radiation \citep{vanPutten:2002ui,vanPutten:2003hd}.

In this work we will discuss and focus on the GW burst 
that will be certainly produced when the core of a super massive 
star collapses to a black hole. This is an intrinsic feature 
of the gravitational collapse to a black hole \citep{Stark1985}.
It is known that the energy emitted in GW through the collapse of a compact object 
to black hole depends strongly on the angular momentum of this object.
Furthermore, the core collapse of a super massive star to 
a black hole is the leading candidate for the central engine 
of a long gamma-ray burst. Thus, every long GRB should be 
accompanied with such a gravitational wave burst. 
This new born black hole with act as 
the central engine and drive the GRB. The electromagnetic energy 
extraction from the rotating black hole sustained through accretion,
is one of the main models (together with accretion onto the black hole) 
to deposit the energy needed to power the burst. 
This mechanism depends strongly on the magnetic field strength and 
topology but the amount of energy that can be extracted is 
purely a result of the angular momentum of the black hole.

Thus, we can tentatively consider that the GW burst and GRB energetics 
both follow  the angular momentum of the collapsing object and the 
resulting black hole. 
If we that the mass distribution of the newborn black holes is narrow and 
mostly concentrated around $\simeq 3 M_{\odot}$, then the above mentioned 
quantities are closely connected and correlated. 
Furthermore, associating the isotropic energy measured 
for a certain GRB,  then $E_{\rm iso}$ depends on the available energy from the black hole which is a function of the dimensionless spin. In what follows we will try to put these ideas 
in a detailed context, describing the physical procedures and estimating 
detectability limits. Moreover, we will highlight the intrinsic 
properties of such bursts while pointing the assumptions that these ideas 
are based on. 

\section{Collapse to a black hole}
\label{sec:nsaid}

In this Section we will review the possibilities of black hole formation,
depending on the mass of the progenitor star and discuss the gravitational wave 
signal that this black hole formation will give. This will be separated 
in the gravitational wave burst produced at collapse and the quasi-normal 
modes that will accompany this burst till the black hole has settled down.
The mass of the progenitor star plays a significant role on the outcome  
of the collapse.
Depending on the mass of the progenitor Wolf-Rayet star (for details see
\cite{Crowther:2006dd}) the procedure that the inner core will 
follow can have a slightly different 
physical path. 
For massive stars, whose mass exceeds $40M_{\odot}$ 
it is expected that the iron core will not
produce an outgoing shock and it will directly collapse to a black hole.
If the mass of the progenitor is in the range between $\sim 23 M_{\odot} 
< M_{\rm prog.} < \sim 40 M_{\odot}$, then the iron core will contract and 
produce a proto-neutron star together with an out-going shock. 
After some time, some of the outward moving stellar matter, which is 
gravitationally bound, will fallback and accrete to the neutron star 
triggering its collapse when it cannot support anymore its own mass.

Both scenarios for the death of massive stars have a common feature, 
the production of a stellar mass black hole and this process 
of gravitational collapse has been thoroughly studied 
and has an intrinsic gravitational imprint, a gravitational burst 
accompanied with quasi-normal modes of the newly born excited black hole 
till it settles down 
\citep{Stark1985,Baiotti:2004wn,Baiotti:2006wm,Baiotti:2007np,Duez:2004uh}. 
The first work on this subject in numerical relativity (in axisymmetry) 
was done by \cite{Stark1985}. 
They computed the first waveforms
and estimated the GW energy from a rotating stellar collapse in axisymmetry.
The 	overall results showed that the 
collapse depends on the dimensionless spin of the initial star. 
The waves are similar but with an increasing amplitude with increasing spin parameter.
A systematic study in gravitational-wave emission from the collapse of neutron stars 
to rotating black holes in three dimensions was done by \citep{Baiotti:2007np}.

They made an analysis on the collapse of stars with different initial 
rotation rates and they confirmed that there exist a precise 
scaling in terms of the dimensionless spin parameter. 
In detail what they found is that the energy radiated in gravitational waves during and after 
the collapse scale as the dimensionless spin in the fourth power 
($\propto (J/M^2)^4$). 
It was reported in \citep{Baiotti:2004wn} that the measured angular momentum of the newly formed black hole after the collapse is remarkably close to the initial one. 
Thus, knowing the spin parameter of the black hole formed will enable us to have an estimate 
on the energy release in gravitational radiation. 
This relation does not scale up to all spin parameters, as it was found that the GW emission from collapsing neutron stars has a maximum in the emitted GW energy. 
This maximum was found to be the result of a delayed collapsed due
to the angular momentum resisting the collapse and thereby
causing a smaller perturbation in the metric fields during
the collapse. 
This correlation of the emitted energy with the spin 
parameter will allow us in the upcoming section to estimate  the energy 
of the gravitational wave burst through the observed isotropic 
luminosity of a long-GRB, which as we will discuss can be also 
related directly to the spin of the newly born black hole. 

They further discuss and compute the overall efficiency 
of converting the binding energy in gravitational waves, which  
is $\delta M /M \simeq 10^{-7}-10^{-6}$ and thus it seems not to be 
an efficient process. This efficiency is calculated by imposing 
to the initial models a pressure reduction to trigger the collapse.
However, it is difficult to know exactly the physical conditions when 
the collapse is triggered and starts. 
The above mentioned efficiency can be viewed as a lower limit to the gravitational wave production.
In the same study by \citep{Baiotti:2007np} another series of collapses 
was performed not by reducing the pressure, but by adding an inward radial 
velocity  $0.02 ~c $ to the interior matter. 
This may not be too far from reality, since velocity perturbations could happen in the interior shells of a massive star which is about to collapse. By imposing these velocity perturbations 
as an initial collapse trigger the efficiency goes up two orders of magnitude ( $\delta M/M \simeq 10^{-5}-10^{-4}$). The energy has the same scaling  
dependence on the dimensionless angular momentum to the fourth power, but 
it is interesting to notice that in this case where the collapse is triggered
by velocity perturbations there is no maximum found, thus this 
energy scaling could go up to high spins. This means that, if we assume that 
the black hole, formed during a long GRB, is highly rotating, then 
the respective GW burst carries significant amount of energy. 
 
   \begin{figure}
	\begin{center}
	  \includegraphics[width=1.0\columnwidth]{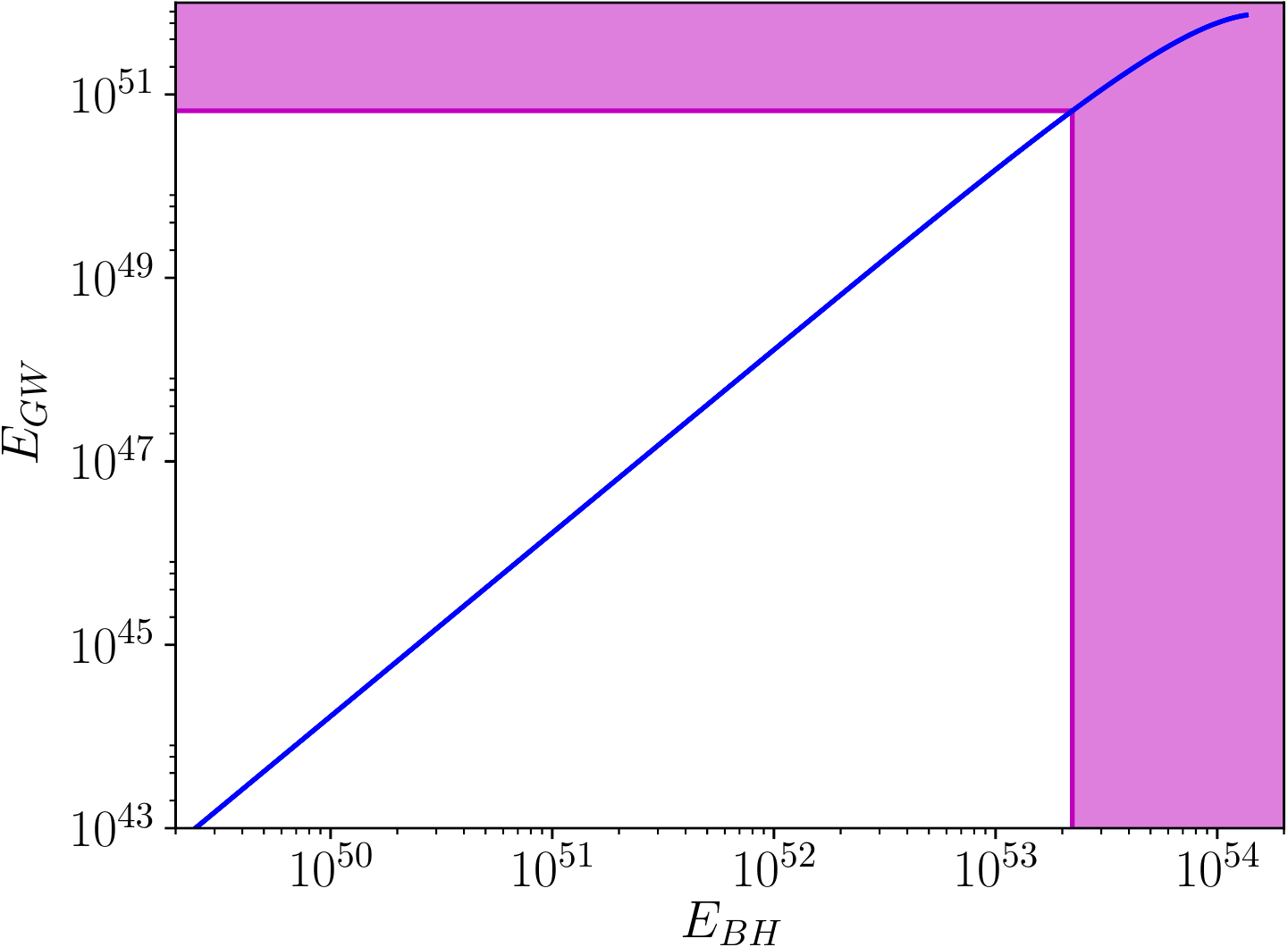}
	\end{center}
	\caption{The available energy from the black hole is plotted 
	versus the energy in gravitational waves, both depend on the 
	dimensionless spin parameter. The line  where the shaded area starts 
	indicates the model with spin parameter $\alpha \simeq 0.54$,  which 
	could be detectable if it  happens closer than 
	$\sim 60 {\rm Mpc}$.
	\label{fig:EGW-EBH}}
\end{figure}
To quantify the energy carried away in GW we use results from their models 
as they report them. The maximum energy emission comes from the most rapidly rotating 
star which gives  $E_{\rm GW} \approx 1.4 \times10^{-6}(M/M_{\odot}) M_{\odot}c^2$ and under 
the assumption that this collapsed $10$kpc  away, they find an upper limit 
for the characteristic strain $h_c$ of $\approx 5.5 \times 10^{-22} (M/M_{\odot})$ 
at a characteristic frequency $f_c$ of $\approx 500 {\rm Hz}$. These limits 
where found for the initial LIGO detectors by convolving their GW signals 
with the detector noise curves. Advanced  LIGO's significant sensitivity 
improvement at low frequencies can see deeper almost one order of magnitude in 
the characteristic strain. We regard the above mentioned limit as a lower limit 
meaning that this was calculated by a pressure reduction, which for an 
isolated collapsing neutron star could be the case, but in the context 
of collapsing shells of a massive star ($12M_{\odot}<M<100M_{\odot}$)
that accrete onto  a proto neutron star which will subsequently 
collapse to a black hole, we believe that velocity perturbations could 
trigger the collapse. Thus, we will estimate in this context the 
characteristic strain by using energy scaling given by \citep{Baiotti:2007np} 
from the velocity perturbation. We also let the order of the strain to 
be $10^{-23}$, which should make the gravitational waves detectable around half a kHz. 
We further assume that the collapsing core to a black hole has mass 
of $\approx 3M_{\odot}$, which is close to maximum mass 
of a differentially rotating neutron star \citep{Baumgarte:1999cq}, of course more details on the stiffness of the equation of state are needed for the precise limit.  
Under these assumptions and requirements we can parametrise the characteristic strain of the GW as follows:
\begin{equation}
h_c \simeq 10^{-23} \left(\frac{E_{\rm GW}}{4.2 \times 10^{-4} M_{\odot}c^2} 
 \right)^{1/2} \left(\frac{r}{33 Mpc}\right)^{-1}\,,
\label{hc}
\end{equation}
which is based upon the values estimated for the most rapidly rotating model with the initial 
radial velocity as the collapse trigger. 
This model has an initial dimensionless spin of $a = 0.54$. 
We can write the energy released in GW as follows:
\begin{equation}
E_{\rm GW}\simeq 1.4\times 10^{-3} \alpha^4 \left( \frac{M_{\rm BH}}{M_{\odot}}\right) M_{\odot} c^2\,,
\label{gw}
\end{equation}
where $\alpha$ is the dimensionless spin parameter. For a spin of  $a = 0.54$ 
would correspond to the estimate of eq.~ \eqref{gw} and for spin of $\alpha = 0.1$ 
would yield to lower point of the right panel of figure $10$ of \citep{Baiotti:2007np}.
 
Here, we can make a comment about the spin of the final black hole related 
to the object's angular momentum just before collapse.
Through accretion  angular momentum  will be transferred to the inner core and reach higher values than what one can expect from an isolated neutron star, in such case 
the resulting black hole could be even close to maximal rotation \citep{Woosley2006ApJ}. 
This means that moments before collapse the compact object has gained angular momentum higher than the breakup limit of a neutron star, and probably could support more mass than 
the usually estimated   maximum mass \citep{Baumgarte:1999cq}. 
If we assume that the relation of the GW energy radiated 
during collapse, which 
depends to the dimensionless angular momentum (to the fourth power), 
could scale up to higher spins, in the case of velocity perturbation trigger,
then these events could be visible to 
advanced LIGO up to a limit of $\sim 100 {\rm Mpc}$.

Since we are discussing about gravitational collapse to black hole we 
should also refer to quasi-normal-modes (QNM). The perturbed black hole horizon 
upon its formation will ring down till it settles down in axisymmetry.
The QNM of a black hole depend on the mass of the black hole and its dimensionless 
angular momentum. There exist thorough studies on black hole QNM 
resulting in approximate formulas with high accuracy ($\simeq 5 \%$). 
The QNM are not expected to be detected, even in the collapse of 
a galactic neutron stars \cite{Ott:2008wt}.
 
\section{Electromagnetic output}
\label{sec:EM}
Let us focus now on the expected electromagnetic radiation from 
the core-collapse of a massive star and the production of a long-GRB. 
We will not make any assumption on the production of high-energy gamma photons, we rather discuss about the reservoir of energy that will then give rise to the GRB.
If the energy input for a long-GRB is given by the newborn black hole, then 
one of the main physical mechanisms is the electromagnetic extraction 
of energy from the spinning black hole \citep{Blandford1977}. 
This mechanism is widely discussed in GRB literature  \citep{Komissarov2009,Nagataki2009}. 
We will give a brief description of this mechanism here. Wolf-Rayet 
stars ending their life are thought to be the progenitors of long GRBs \citep{Woosley:2006fn}. It is observed that Wolf-Rayet stars are magnetized \citep{delaChevrotire2014}. 
During the collapse, advected matter will bring the preexisting magnetic flux 
at the vicinity of the black hole. This magnetic flux is trapped by the the outer shells of the   collapsing star, which will act as a massive disk  to keep all these magnetic flux $\Psi$ close to the horizon. 
This context is enough to give rise to a powerful Poynting dominated jet, 
which will be contaminated with baryons while  drilling through the star. 
The rate that the black hole is losing energy is 
$\displaystyle{\dot{E} \approx -\frac{1}{6\pi^2 c}\Psi^2\Omega^2}$, where $\Omega$ is the angular 
velocity of the black hole horizon, a function of the spin parameter. This is 
a widely studied and confirmed mechanism, both analytically and numerically  
\citep{Komissarov2001,Mckinney2006, Komissarov2007, Nathanail2014, Gralla2016}.  
According to \cite{Christodoulou1971} the available rotational energy for a  
black hole depends on its dimensionless spin parameter ($a= J/M^2$) and reads as: 

\begin{equation}
E_{\rm rot} = \left[ 1- \frac{1}{2} \sqrt{(1+\sqrt{1 - a^2})^2 +a^2}\,\right] \times M_{\rm BH} ~ c^2\,
\label{Erot}
\end{equation}
To give an idea for the available energy for a black hole of $\simeq 3 M_{\odot}$ we can give some  values including two extreme values, for a spin parameter of $a= 0.008$, which we could say that  is almost a non-rotating (Schwarzschild) black hole, nevertheless  the available energy to extract is $\simeq 4 \times 10^{49} {\rm erg}$,  whereas for a rotating black hole close to maximal rotation with spin $a= 0.98$, the available energy is $\simeq 1.25\times 10^{54} {\rm ergs}$. 
In order for these huge amounts of energy to produce a GRB, a highly relativistic 
collimated plasma erupting  at the outer shells of the star is needed. 
 
This energy budget will be converted to radiation and will show up in the 
gamma-rays (details on the huge literature in \cite{Kumar2015}). From 
the observed high-energy radiation an estimate on the overall 
energetics can be made. If the distance to a GRB is known then the energy released 
can be estimated, thus for GRBs with known redshift the energy 
can be calculated assuming the burst is isotropic. For GRBs  detected by 
Swift and GBM (onboard FERMI) the isotropic energy spans more than 
six orders of magnitude from $10^{48} {\rm erg}$ till close to $10^{55} {\rm erg}$ \citep{Cenko2010,Fong:2015oha}. 
However, it is believed that these explosions are not isotropic and they are actually 
beamed  resulting from a relativistic jet emerging in the vicinity of the 
black hole and drilling through the star \citep{Zhang:2003rp} 
to make itself visible  when reaching a photosphere \citep{Rees2005}. 
This energetic bursts give extreme luminosities, measured at the peak of the 
lightcurve, the isotropic luminosity in gamma-rays ranges 
between $\simeq 10^{47} {\rm erg s^{-1}}$ and $\simeq 10^{54} {\rm erg s^{-1}}$ \citep{Kumar2015}. 
This wide range in luminosities could be due to the difference 
in the magnetic field properties and strength close to the vicinity 
of the central engine,which is  dictated by the above mentioned 
electromagnetic mechanism.
 
In order to relate the intrinsic energy that comes straight from 
the spin of the black hole to the observed energy, we should also take into account 
the beaming factor. 
To do that we should have an estimate on the 
opening angle of a GRB jet. 
By modeling the GRB afterglows in different energies and 
wavelengths (X-rays , radio) the jet break can be pinpointed. 
Opening angles have been observed to be in the tide range of 
$2 - 10 $ degrees \citep{Rhoads1999,Sari1999}. 
Assuming that the relativistic jet is conical would yield a solid angle of 
$\Omega_{\gamma} = 2 \pi ( 1- cos\theta)$. 
The isotropic energy calculated should be corrected as follows to give the 
true amount of energy released $\displaystyle{E_{\rm iso}\simeq \frac{\Omega_{\gamma}}{4 \pi}}$, from this correction the true range of energies can be accounted. 
The small range for opening angles observed, allow us to take the same  correcting 
factor  for all cases.  
The isotropic energy should be corrected by a factor of $5\times 10^{-3}$ 
(this is the exact case for an opening angle of $8$ degrees, assuming a conical jet).
As a result, the above mentioned observed and estimated isotropic energies 
give a  range between $ 5 \times 10^{48} {\rm erg} $ and $ 5 \times 10^{52} {\rm erg}$.
We have to state here that the discussion for the opening angle is limited to 
the brightest GRBs for which broadband observations exist. As pointed out by 
\citep{Cenko2010} it would not be surprising if most Swift events have larger 
opening angles (or even isotropic), which would make jet-break estimates 
and measurements more difficult to account \citep{Perna2003}. 

\begin{figure*}
\begin{center}
\includegraphics[width=1.0\textwidth]{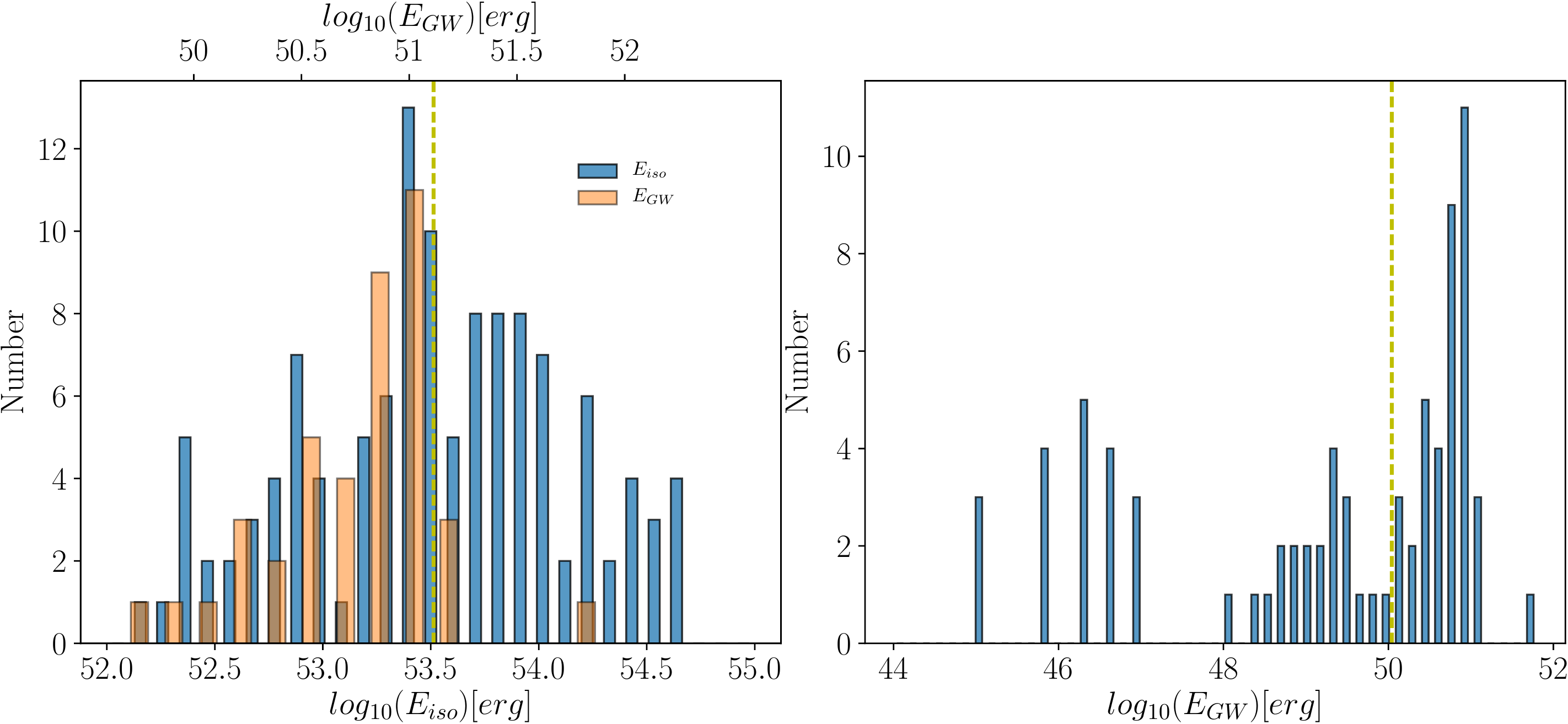}
\end{center}
\caption{Histogram showing the distribution of $E_{iso}$ for GRBs with known redshift observed by Fermi/GBM and Konus Wind, and the distribution of the estimated energy released in GW. The $E_{\rm iso}$ is from observed data, whereas the $E_{\rm GW}$ data are estimated using black hole spin parameters  from simulation results for progenitors of long-GRBs. The yellow dashed line shows the mean value of the $E_{\rm iso}$ (left panel) and $E_{\rm GW}$ (right panel) respectively. 
On the left panel we overplot  the high energy part of the $E_{\rm GW}$ distribution that 
corresponds to the values of $E_{\rm iso}$.\label{Histo}}
\end{figure*}  
We will tentatively try to  relate  every burst  that the
observable isotropic 
energy can be estimated (e.g. with known redshift), to a certain spin
parameter for its black hole powering the burst. Having discussed all the above, 
we decided to drop any assumption for the beamed luminosity 
coming towards us. This is because, assuming that for all bursts 
the correcting factor is lying inside a tide range, the distribution 
of the isotropic energy can be viewed as a distribution in spin parameters.
The only difference would be an overall shift of two orders of magnitude, 
but the scaling will be the same.

In this respect we can invert equation \eqref{Erot} 
\begin{equation}
a = 2\times \sqrt{2-c}(-1+c)\sqrt{c}\,
\label{spin}
\end{equation}
where $c=f \times E_{iso}/M_{\rm BH} c^2$ and $f $ 
is a factor correcting the isotropic energy in terms of beaming 
and also taking into account the efficiency of the radiation process. 
We can assume that $f $ is a universal constant, in other words 
for all GRBs it remains almost the same. Then we can view the distribution 
of GRB isotropic energy as an identical one of the distribution of  
spins of the black holes which act as the central engines of the bursts. 
We could tentatively have as a starting point the higher energetic ones, 
by limiting the higher possible spin from theory \citep{Dessart2012}.   

We should also make a point here, that the spin of the black hole 
may slightly change during the course of the GRB. This can be accounted  
to the  influence from the accretion disk that will form around it and 
also to episodic mass infall as big parts of the collapsing star enter the 
horizon, which could also give rise to  energetic X-ray flaring activity 
\citep{Nathanail:2015tga}. Of course, all this flaring activity and the structure 
of the X-ray afterglow is less energetic that the gamma-rays.   
The reason we can associate the isotropic 
energy to the original spin of the black hole is that  
the isotropic energy is governed by the prompt gamma-ray emission.

We have discussed so far how, from the isotropic energy of long-GRBs 
with known redshift, 
we can account for the spin parameter of  newly born black holes after 
the core-collapse of super massive stars. The next step is to correlate  
the energy release after the formation of the black hole with 
the energy released in gravitational waves just before the GRB trigger, 
during the collapse to a black hole. As we discussed above, our knowledge from numerical studies of the  gravitational collapse to a black hole is that 
the spin parameter 
of the resulting black hole is close to the spin (dimensionless angular momentum) 
of the initial compact star (or compact core) 
before collapse \citep{Baiotti:2007np}. 
Thus, having in hand the estimate for the spin, from the isotropic energy 
output, we can estimate the intensity of the gravitational wave burst 
produced during the collapse to a black hole. Of course,  based on the current 
threshold limits of the interferometric detectors and the most optimistic 
expectations for the  energy  of such bursts, the closest would be  
$\sim 100{\rm Mpc}$, if we expect a simultaneous  detection from such an explosion. 
Nevertheless, we should state  that the closest long-GRB detected is 
GRB980425 at a redshift of $z=0.008$ 
\citep{Tinney1998, Soffitta1998}
which is at $\sim 34 {\rm Mpc}$,  
which is closer than 
the estimated distance  for a gravitational wave burst 
detection during  the birth of a long-GRB. 

The conclusion of all the above discussion, is that the energy released 
in GW during the collapse of the core of a massive star 
is related to the energy release in electromagnetic radiation, this is 
straight forward since both depend strongly on the spin parameter 
of the black hole. For the electromagnetic radiation this can be regarded 
as an assumption since other mechanisms have also been invoked 
to power the bursts. 
In Fig. ~\ref{fig:EGW-EBH} we show the relation between the available energy
from the black hole and the corresponding energy that would be released 
during the gravitational collapse of the central compact core 
of a massive star to a black hole. The shaded area (colored magenta) 
is the region that could be possible detected by present interferometric 
detectors, the limit being the model discussed above with a spin parameter 
of $\alpha \simeq 0.54$. 

Some models in the shaded area could be detected as far as  $\sim 100 {\rm Mpc}$.
It is clear from eq. ~\eqref{hc} that, 
for every order of magnitude increase in 
sensitivity, we gain an order of magnitude in distance further 
away. Some points that have to be made about this relation are in order here. 
The energy released in gravitational waves can be regarded as an upper limit 
since they were taken under the assumption that the collapse is triggered 
by velocity perturbations. For an isolated neutron star this is not expected to 
happen, but in the context of accreting  matter onto a proto-neutron star 
 in the core of a massive star this could be a possibility. 
A lower limit for the production of the GW burst could be given by the values given when the collapse is triggered by pressure reduction. This would scale down the relation in Fig.~ \ref{fig:EGW-EBH}, $1-2$ orders of magnitude, making them visible only as close as 
$\sim 1 {\rm Mpc}$. 
Another comment has to be made about the existence of an intrinsic maximum on the 
gravitational energy release during the collapse. In the case of the 
pressure reduction trigger,   its existence is clear. 
But in the case of velocity perturbation trigger, 
this haven't appeared in the range of spins used, 
with the higher spin being around $\alpha \simeq 0.6$ 
\citep{Baiotti:2007np}.  

Furthermore, higher rotation for an initial neutron star 
would be close to the break up limit. However, in the context of a progenitor  
of long GRB it is expected that during the death of a massive star the inner regions 
of the star possess large amounts of angular momentum, maybe two orders of magnitude 
larger than the ones that give birth to ordinary neutron stars. Some of these would 
produce a black hole close to maximal spinning \citep{Woosley2006ApJ}. 
On the other hand, other studies have shown that 
the condition of initial rapid rotation 
for the massive star could possibly not give birth to highly spinning black holes 
due to the production of a magneto-rotational explosion that would prevent 
black hole formation. The spin parameter range they found for 
black holes formed after the death of fast rotating massive stars is bound 
$\alpha < \simeq 0.6$ \citep{Dessart2012}. 
If we translate this value in an upper bound for the available 
energy of the black hole, this is close to $10^{53}$ which is 
similar to the upper limit for the beaming-corrected energies from GRBs.
Overall this is in favor of such a central engine.
  
Following the correlation discussed so far, we want to discuss how it
would show up in both the distribution of the $E_{\rm iso}$ and $E_{\rm GW}$, 
once both have been observed. Since, 
for the time being no such gravitational wave burst
detections exist, we use simulation data from core-collapse of massive stars. 
The evolution of rapidly rotating massive stars has been studied, in order to show the possibility of reaching the requirements needed for a GRB to be launched, when the black hole is formed. Results from these studies  have estimated the value of the spin parameter. We use the results from tables 1 and 2 of \cite{Woosley2006ApJ} and Table $1$ of \cite{Dessart2012}. 
Following the discussion in \cite{Dessart2012}, we have left all models 
that produce a  maximally rotating black hole out of this study. Using all 
the spin parameters for these models that produce a rotating black hole 
and could supposedly be progenitors of long-GRB, we apply 
eq.~\eqref{gw} to find the distribution of GW energy from such bursts.
(right panel of Fig.~\ref{Histo}). The inclusion of very small 
spin parameters make the distribution span almost $8$ orders of magnitude and 
the mean value (yellow dashed) not aligned with the main peak. To produce 
the corresponding distribution of the isotropic energy of GRBs with known 
redshift we used the data from Table $2$ and $3$ of \cite{Atteia2017}. 
These include $52$ GRBs detected by Fermi/GBM and another $69$ by Konus-Wind.
Their distribution is shown in blue on the left panel of Fig.~\ref{Histo}.
To see how the gravitational wave energy trace 
this distribution, we overplot on the 
same panel the far right part of the distribution of the GW energy (energies 
between $10^{49.5}$ and $10^{52}$) shifted to the corresponding 
observed isotropic energies. 
As a first result these simulated data seem to trace well the observed distribution 
of $E_{\rm iso}$. The cutoff after 
the peak in the gravitational waves energy comes from the fact that in 
\cite{Dessart2012} the spin of black holes formed  is bound by $\alpha \sim 0.6$.

\section{Conclusion}
\label{sec:conc}

The gravitational collapse of the central compact core 
of a massive star to a black hole is thought to be the 
central engine of long GRBs. This gravitational collapse 
will give rise to a GW burst, which 
release energy proportional to its dimensionless angular momentum. 
The available rotational energy of a spinning black hole also depends 
of the dimensionless spin parameter. Thus, the GRB isotropic energy 
can give an estimate on the underlying spin of the black hole.
Then this can be used in order to estimate the energy released 
in GW. 

A thorough automatic search could be established in the data of 
interferometric detectors after the 
electromagnetic detection of a GRB in order to find 
such GW bursts at the time just before this GRB was detected. 
In the future, if the sensitivity of the present detectors 
is enhanced $1-2$ orders of magnitude, then these bursts could 
be observed as far as $1 - 10 {\rm Gpc}$,

\section*{Acknowledgements}
It is a pleasure to thank C.Fromm, O. Porth and L. Rezzolla 
for useful discussion.
A.\,N.\, is supported by an Alexander von Humboldt Fellowship.
Partial support comes  from ``NewCompStar'', COST Action
MP1304.  
M.\,D.\,L.\ is supported by the ERC Synergy Grant
``BlackHoleCam'' -- Imaging the Event Horizon of Black Holes (Grant
No.~610058). 
M.D.L. acknowledge COST Action CA1511 Cosmology and Astrophysics 
Network for Theoretical Advances and Training Actions (CANTATA), 
supported by COST (European Cooperation in Science and Technology) and INFN Sez. di Napoli (Iniziative Specifiche QGSKY and TEONGRAV).

\section*{}
\bibliographystyle{mnras}

\label{lastpage}
\end{document}